# Collapse suppression and stabilization of dipole solitons in two-dimensional media with anisotropic semi-local nonlinearity

Fangwei Ye, [1*] Boris A. Malomed, [2] Yingji He [1] and Bambi Hu[1, 3]

[1]*Department of Physics, Centre for Nonlinear Studies, and The Beijing-Hong Kong Singapore Joint Centre for Nonlinear and Complex Systems, Hong Kong Baptist University, Kowloon Tong, China*

[2]*Department of Physical Electronics, School of Electrical Engineering, Faculty of Engineering, Tel Aviv University, Tel Aviv 69978, Israel*

[3]*Department of Physics, University of Houston, Houston, Texas 77204-5005, USA*

*Corresponding author: fwye@hkbu.edu.hk*

We consider the impact of anisotropic nonlocality on the arrest of the collapse and stabilization of dipole-mode (DM) solitons in two-dimensional (2D) models of optical media with the diffusive nonlinearity. The nonlocal nonlinearity is made anisotropic through elliptic diffusivity. The medium becomes semi-local in the limit case of 1D diffusivity. Families of fundamental and DM solitons are found by means of the variational approximation (VA) and in a numerical form. We demonstrate that the collapse of 2D beams is arrested even in the semi-local system. The anisotropic nonlocality readily stabilizes the DM solitons, which are completely unstable in the isotropic medium.

## 1. Introduction

The spatial nonlocality of the nonlinear response of transparent materials to propagating electromagnetic waves implies that the nonlinear part of the polarization of the medium is determined not only by the wave's intensity at the same point, but also by the distribution of the intensity around it [1, 2]. This situation occurs when the nonlinearity involves the diffusion of carriers or long-range interactions, in media such as liquid crystals featuring the long-range re-orientational nonlinearity [3,4], vapors where the atomic diffusion causes the transport of excitations away from the region of the light-matter interaction [5,6], and materials with the thermal mechanism of the nonlinearity generation, where weak absorption of light causes local heating, smeared by the heat conductivity, and the corresponding small variation of the local refractive index, in proportion to the temperature perturbation [7,8]. The nonlocal nonlinearity is also characteristic to Bose-Einstein condensates of atoms carrying magnetic moments, which give rise to the long-range anisotropic dipole-dipole interactions [9]. The spatial nonlocality may significantly change the shape and dynamics of light beams, leading to such effects as the arrest of the collapse of the beams, and stabilization of otherwise unstable complex species of spatial solitons. The collapse, i.e., formation of a field singularity after a finite propagation distance, is inherent to multidimensional beams in local Kerr media [10-12]. However, using several particular forms of the nonlinear-response kernels, it was first explicitly shown in Refs. [13, 14] that the collapse may be suppressed in nonlocal media. Later, more general kernels were used to prove the elimination of the collapse by the nonlocality [15, 16].

Another essential consequence of the nonlocality is that it can change interactions between solitons in the nonlinear medium. It is well known that, in local nonlinear materials, out-of-phase bright soliton always repel each other, hence DM solitons (which seem as bound states of two fundamental solitons with opposite signs) may only exist in the form of vector states in two-component settings, in which the fundamental soliton created in one component supports a stable DM counterpart in the other [17-20]. An alternative approach to the creation of DM complexes is the use of an underlying periodic (lattice) potential, where the interplay of the repulsion between the two "monopoles" and their pinning by the lattice gives rise to stable bound states, in one-dimensional (1D) [21] and two-dimensional (2D) [22] cases alike. In contrast to that, in nonlocal nonlinear media, out-of-phase bright solitons can *attract* each other, and may thus form bound states, i.e., *scalar DMs* [23-33]. Experimentally, stationary 2D *multipole-mode* solitons have been observed in media with the thermal nonlinearity [23]. Nevertheless, all stationary *non-rotating* multipoles, that were thus far created experimentally in nonlocal materials, are subject to a weak instability, therefore attention was naturally drawn to the problem of stabilization of such modes. In this direction, it has been predicted that, in liquid crystals, dipoles may become stable in a very narrow parametric region, at low powers [24]. In very recent studies, the stabilization of DMs was demonstrated in the model of media with the thermal (diffusive) nonlinearity set up in a rectangular 2D domain, as well as their stabilization by saturation of the nonlocal nonlinear response [25]. It was also reported that stable DMs are possible in a more complex model of the atomic-vapor optical medium, which involves both the nonlocality and saturation [6]. It should be noted that the response functions (kernels accounting for the nonlocality), considered in the above-mentioned works, including the investigation of the suppression of the collapse and stabilization of the dipoles, were all isotropic. However, nonlocality may be anisotropic in the available materials. Actually, the study of effects of the anisotropy of the nonlocality remains an open problem.

Our purpose in this work is two-fold. First, we study the influence of the anisotropic nonlocality on the collapse. While it is natural to expect that the collapse remains arrested by the nonlocality in the general anisotropic case, we find that even in the limit case of the *one-dimensional* nonlocality in the 2D setting, the medium, which remains local in one direction, is able to completely eliminate the collapse. Second, we demonstrate that DMs, which were completely unstable in the isotropic nonlocal medium, become *stable* under the action of the anisotropic nonlocality, provided that the total power exceeds some critical value. Further, the stability region for the DM solitons significantly expands with the increase of the anisotropy.

**2. The model and the variational approximation**

As a natural anisotropic generalization of the common model with the nonlocal cubic self-focusing nonlinearity [3, 4, 16], we adopt the following system of scaled equations for complex amplitude $q$ of the electromagnetic wave (assuming a single polarization of light) and local correction $n$ to the refractive index,

$$i\frac{\partial q}{\partial z} = -\frac{1}{2}\left(\frac{\partial^2 q}{\partial x^2} + \frac{\partial^2 q}{\partial y^2}\right) - qn, \tag{1a}$$

$$n - d\left[(1+\varepsilon)\frac{\partial^2 n}{\partial x^2} + (1-\varepsilon)\frac{\partial^2 n}{\partial y^2}\right] = |q|^2. \tag{1b}$$

Here $z$ and $(x,y)$ are the longitudinal and transverse coordinates scaled to the characteristic beam's width, $r_0$, and the respective diffraction length, $L_{\text{diff}} = k_0 r_0^2$, respectively, while $\sqrt{d}$ is the correlation length of the nonlocal response. By means of an obvious additional rescaling, which is admitted by Eqs. (1), we may actual set $d \equiv 1$. Parameter $\varepsilon$ ($0 \leq \varepsilon \leq 1$) in Eq. (1b) is an effective eccentricity (the anisotropy parameter). While $\varepsilon = 0$ in the isotropic medium, $\varepsilon = 1$ corresponds to the limit case of a *semi-local* material, with the nonlocality acting only in the $x$ direction, whereas in the $y$ direction the nonlinear response is completely local. Equations. (1) conserves the total power, $U = \iint_{-\infty}^{\infty} |q|^2 dxdy$, along with the corresponding Hamiltonian and two components of the momentum. Note that, following the commonly adopted assumption [3-8,15,16,23-33,38,39], in Eq. (1b) we neglect the heat diffusivity, and related nonlocality, in the longitudinal direction, because the group velocity of light is much larger than the velocity of the propagation of the heat wave (this may be different, in principle, in specific media admitting ultraslow light transmission, see, e.g., Ref. [34]).

The Laplacian in Eq. (1a) represents the diffraction of the optical beam (in the paraxial approximation), hence it is always isotropic, as it does not depend on a particular material (taking the same form in the absence of any material). On the other hand, in the case of the most relevant thermal (diffusive) mechanism of the nonlinear response, the spatial operator in Eq. (1b) represents the diffusivity in the medium, which may easily be made anisotropic. In particular, in materials resembling "1D metals", which are composed of long parallel threads (typically, these are polymer chains), the heat diffusivity may become effectively one-dimensional acting only along the chains and being suppressed in the transverse directions by gaps separating the chains, which corresponds to the limit case of $\varepsilon = 1$ in Eq. (1b) (see, e.g., Ref. [35]). In the latter case, the light propagation direction is perpendicular to the orientation of the material-forming chains.

Equations (1) bear some formal similarity to the above-mentioned two-component models that admit the stabilization of DMs in one component by a fundamental soliton in the other. Indeed, in the present case the DM structure will be demonstrated in the field variable, *q*, while the coupled refractive-index perturbation is shaped more as a fundamental soliton – in the sense that it does not change its sign in the center – see, e.g., Fig. 2 below, and respective *ansätz*e represented by Eqs. (4b) and (5). However, the qualitative difference in the structure of the present system from that governing the co-propagation of the two interacting fields is that, in Eq. (1b), $|q|^2$ is the source generating $n(x,y)$, while in the coupled propagation equations (nonlinear Schrödinger equations) the nonlinear interaction between the two fields is of the XPM (cross-phase-modulation) type.

Stationary solutions to Eqs. (1) with propagation constant $b$ are sought for as $q = f(x,y)\exp(ibz)$, $n = n(x,y)$, where real functions $f(x,y)$ and $n(x,y)$ obey the following equations:

$$\frac{1}{2}\left(\frac{\partial^2 f}{\partial x^2} + \frac{\partial^2 f}{\partial y^2}\right) + nf = bf,$$

$$n - d\left[(1+\varepsilon)\frac{\partial^2 n}{\partial x^2} + (1-\varepsilon)\frac{\partial^2 n}{\partial y^2}\right] - |f|^2 = 0, \qquad (2)$$

which can be derived from a Lagrangian,

$$L = \frac{1}{2}\iint\left\{\left(|f_x|^2 + |f_y|^2\right) + 2b|f|^2 - 2n|f|^2 + d\left[(1+\varepsilon)n_x^2 + (1-\varepsilon)n_y^2\right] + n^2\right\}dxdy, \qquad (3)$$

The variational approximation (VA) may be used to predict fundamental and DM solitons in the present model (the VA was applied to the isotropic model in Ref. [16]), using the following natural anisotropic *ansätze* for the fundamental and DM solitons, respectively,

$$f_{\text{fun}} = A\exp\left[-(1/2)\left(\alpha x^2 + \beta y^2\right)\right], \qquad (4a)$$

$$f_{\text{dip}} = Ax\exp\left[-(1/2)\left(\alpha x^2 + \beta y^2\right)\right]; \qquad (4b)$$

in either case, the *ansatz* for the nonlinear refractive index is

$$n = B\exp\left[-(1/2)\left(\gamma x^2 + \delta y^2\right)\right]. \qquad (5)$$

According to these expressions, the solitons' widths along $x$ and $y$ are given by, respectively, $w_x = 1/2\sqrt{\alpha}$, $w_y = 1/2\sqrt{\beta}$, and the integral powers of the fundamental and DM solitons are $U_{\text{fun}} = \pi A^2 / \sqrt{\alpha\beta}$ and $U_{\text{dip}} = \pi A^2 / \left(2\alpha\sqrt{\alpha\beta}\right)$.

Generally, one may expect the existence of two different species of DMs, with mutually perpendicular orientations. We here focus on the single type, aligned with the nonlocal direction [$x$, in the present notation – see Fig. 2(c) below], as the DM oriented in the orthogonal direction cannot exist in the semi-local ($\varepsilon = 1$) limit, due to the repulsion between its constituents.

Further, setting $\varepsilon = 1$, substituting expressions (4) and (5) into Eq. (3), and performing the spatial integration, we arrive at the following effective Lagrangians for the fundamental and DM solitons:

$$\frac{L_{\text{eff}}^{(\text{fun})}}{\pi} = \frac{A^2}{4}\left(\sqrt{\frac{\alpha}{\beta}} + \sqrt{\frac{\beta}{\alpha}}\right) - \frac{2BA^2}{\sqrt{(\gamma+2\alpha)(\delta+2\beta)}} + \frac{bA^2}{\sqrt{\alpha\beta}} + \frac{B^2}{2\sqrt{\gamma\delta}} + \frac{dB^2}{2}\sqrt{\frac{\gamma}{\delta}}, \quad (6)$$

$$\frac{L_{\text{eff}}^{(\text{dip})}}{\pi} = \frac{3A^2}{4\sqrt{\alpha\beta}} + \frac{A^2\beta}{4\alpha\sqrt{\alpha\beta}} + \frac{bA^2}{\alpha\sqrt{\alpha\beta}} - \frac{4BA^2}{\sqrt{(\gamma+2\alpha)(\delta+2\beta)}(\gamma+2\alpha)} + \frac{B^2(d\gamma+1)}{\sqrt{\gamma\delta}}. \quad (7)$$

Then, from a straightforward semi-analytical/semi-numerical solution of the corresponding Euler-Lagrange equations, $\partial L_{\text{eff}} / \partial(A, B) = \partial L_{\text{eff}} / \partial(\alpha, \beta, \gamma, \delta) = 0$, we find parameters $(A, B, \alpha, \beta, \gamma, \delta)$ as functions of $b$ and $d$. In particular, for the fundamental soliton we obtain $A^2 = \frac{2dB^2(\alpha/\beta)^2}{1-(\alpha/\beta)^2}$, which implies $\alpha < \beta$, i.e., the shape of the soliton is elliptic with a longer axis aligned with the nonlocal direction. The results generated by the VA for fundamental and DM solitons are presented in Fig. 1(a) and 1(b), respectively. As expected, the solitons always feature the elliptic shape, with $w_x > w_y$. We also note that, in the entire region of values of the propagation constant, the *Vakhitov-Kolokolov* (VK) criterion [11] holds, $dU/db > 0$, suggesting that all the solitons in the semi-local model should be stable again the collapse, which is confirmed by full numerical results, as shown below.

### 3. Numerical results

To find numerical solutions to stationary equations (2), we used the standard relaxation method. In agreement with the prediction of the VA, the so generated soliton solutions show elliptic shapes at $\varepsilon \neq 0$. Figure 2 displays the soliton profiles in the limit case of the semi-local system, with $\varepsilon = 1$. Due to the nonlocality in the $x$ direction, the induced perturbation of the refractive index extends far beyond the region occupied by the optical field along $x$; in contrast, the local response along $y$ restricts the index perturbation in this direction to the same area which carries the optical density. To quantify the ellipticity of the fundamental soliton, we define its integral widths along $x$ and $y$ as $w_x = \left( \iint x^2 f^2(x,y) dx dy / U \right)^{1/2}$ and $w_y = \left( \iint y^2 f^2(x,y) dx dy / U \right)^{1/2}$. Dependences of the widths and the total power on the propagation constant are displayed in Fig. 1, which demonstrates good agreement of the VA with the numerical results.

An issue of special interest is the effect of the anisotropy of the nonlocality on the instability of localized patterns against the onset of the collapse. Previous studies on the collapse arrest by the nonlocality were performed for isotropic kernels. One may expect that, in the case of the weak or moderate anisotropy, the nonlocality should also readily suppress the collapse. We have found that this is true, running systematic direct simulations of Eqs. (1) with various input conditions, using a symmetrized split-step Fourier scheme. The typical propagation distance was $z = 1000$, with much longer distances used in selected cases.

The most interesting case, when the arrest of the collapse is not obvious *a priori*, is that of the semi-local system, with $\varepsilon = 1$ in Eqs. (1). We have checked this case too by means of the systematic direct simulations of Eqs. (1), using arbitrary inputs or perturbed solitons as initial conditions. Figure 3 presents a typical output of such simulations. In the course of the propagation, the amplitude and widths feature periodic oscillations, see Figs. 3(a,b), without any trend to the collapse.

Thus, a general conclusion is that the 1D nonlocality is sufficient for the *complete stabilization* of the 2D system against the collapse. In this connection, it is relevant to mention that a 1D lattice potential (periodic in one dimension and uniform in the other) may stabilize 2D solitons in local models [36,37]; however, in that case the collapse is not eliminated completely (a strongly compressed soliton will undergo the collapse).

Another essential result that the systematic simulations of Eqs. (1) reveal is the stabilization of the DM solitons in the presence of the anisotropic nonlocality, which would be *unstable* in the isotropic nonlocal model. To address the impact of the anisotropy on the DM stability, we have performed comprehensive simulations of Eqs. (1) with the input in the form of $q|_{z=0} = f_{\text{dip}}(1+\rho)$, where $f_{\text{dip}}$ is the stationary DM solution, and $\rho$ represents broadband noise with variance $\sigma^2_{\text{noise}} = 0.01$. The output of the simulations is that, while the DM solitons are indeed always unstable in the isotropic system, they become stable in the anisotropic model, provided that the propagation constant (hence, the total power too) exceeds some critical value: $b > b_{\text{cr}}$, or $U > U_{\text{cr}}$. The respective *stability regions*, which are plotted in Fig. 4, in parametric planes $(b, \varepsilon)$ and $(U,\varepsilon)$, expand with the increase of the anisotropy. Thus, the stability region disappears at $\varepsilon=0$, and attains the maximum size at $\varepsilon=1$. DM solitons belonging to the stability region retain their initial shapes in the course of indefinitely long propagation, even in the presence of considerable initial random perturbations, as clearly seen in Figs. 5(a) and 5(b). In contrast to that, unstable DMs spontaneously transform themselves into stable fundamental solitons, as Figs. 5(c) and (d) show.

Finally, we mention that other types of complex localized modes, such as quadrupoles and solitary vortices, can also be supported by the anisotropic nonlinearity. However, we have found that the quadrupoles are always unstable, as in the isotropic model [25], while the vortices are actually stable in a narrower region than in the isotropic model, cf. Refs. [38,39] which also demonstrated that strong anisotropy may destabilize solitary vortices, while fundamental solitons are completely stable.

## 4. Conclusion

We have introduced the anisotropic 2D model of nonlinear optical media of the diffusive (thermal) type, and the semi-local system, as its limit form. The model combines the isotropic paraxial diffraction of the optical wave and anisotropic diffusivity (the model becomes semi-local in the limit of the 1D diffusivity). Families of fundamental and DM (dipole-mode) solitons have been found by means of the variational approximation and in the numerical form. We have studied the impact of the anisotropic nonlocality on the suppression of the collapse and stabilization of DMs, concluding that the collapse is fully arrested even in the semi-local limit, and the anisotropy of the nonlocality readily stabilizes the DMs which are completely unstable in the standard isotropic nonlocal model.

# References with titles

# References without titles

**FIGURE CAPTIONS**

FIG. 1. (Color online) The soliton's widths, $w_x$, $w_y$, and the total power, $U$, versus the propagation constant for the families of fundamental (left) and dipole-model (right) solitons in the semi-local system, with $d = \varepsilon = 1$. The variational and numerical results are plotted by continuous and dashed curves, respectively. All quantities are plotted in arbitrary dimensionless units.

FIG. 2. (Color online) Profiles of the absolute values of the field and local refractive-index perturbation, $|q(x, y)|$ and $n(x, y)$ ( left and right columns, respectively), for typical examples of the fundamental and dipole-mode solitons (top and bottom rows, severally) in the semi-local system, with $b = d = \varepsilon = 1$. All quantities are plotted in arbitrary dimensionless units.

FIG.3. (Color online) Oscillations of the amplitude (left) and widths (right) in direct simulations of the semi-local system with initial conditions in the form of a perturbed soliton, $q(x, y)|_{z=0} = f(x, y)(1 + 0.1)$ where $f(x, y)$ is the fundamental soliton at $d = \varepsilon = 1$. All quantities are plotted in arbitrary dimensionless units.

FIG.4 The dipole-mode solitons are stable above the curves in the two parameter planes. All quantities are plotted in arbitrary dimensionless units.

FIG.5 (Color online) Typical examples of the perturbed propagation of stable [columns (a) and (b)] and unstable [columns (c) and (d)] DM solitons, shown in terms of profiles of the absolute value of the field, $|q(x,y)|$ (top row), and the associated nonlinear refractive index, $n(x,y)$ (bottom row). In (a) and (b), $b=3, \varepsilon=1$; in (c) and (d), $b=2, \varepsilon=1$. All quantities are plotted in arbitrary dimensionless units.

**FIGURES**

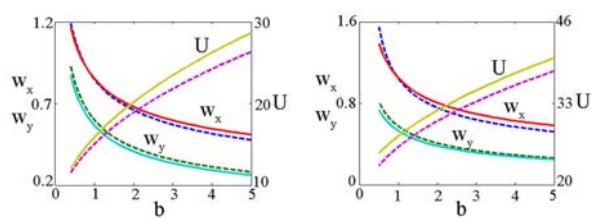

FIG. 1

FIG. 1. (Color online) The soliton's widths, $w_x$, $w_y$, and the total power, $U$, versus the propagation constant for the families of fundamental (left) and dipole-model (right) solitons in the semi-local system, with $d = \varepsilon = 1$. The variational and numerical results are plotted by continuous and dashed curves, respectively. All quantities are plotted in arbitrary dimensionless units.

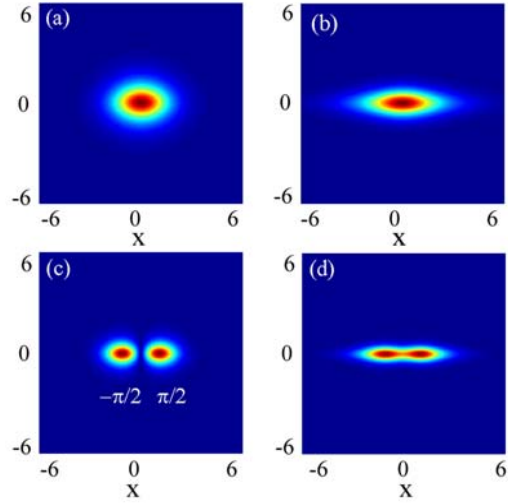

FIG.2

FIG. 2. (Color online) Profiles of the absolute values of the field and local refractive-index perturbation, $|q(x,y)|$ and $n(x,y)$ ( left and right columns, respectively), for typical examples of the fundamental and dipole-mode solitons (top and bottom rows, severally) in the semi-local system, with $b = d = \varepsilon = 1$. All quantities are plotted in arbitrary dimensionless units.

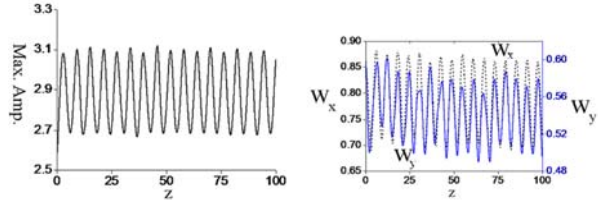

FIG.3

FIG.3. (Color online) Oscillations of the amplitude (left) and widths (right) in direct simulations of the semi-local system with initial conditions in the form of a perturbed soliton, $q(x,y)|_{z=0} = f(x,y)(1+0.1)$ where $f(x,y)$ is the fundamental soliton at $d = \varepsilon = 1$. All quantities are plotted in arbitrary dimensionless units.

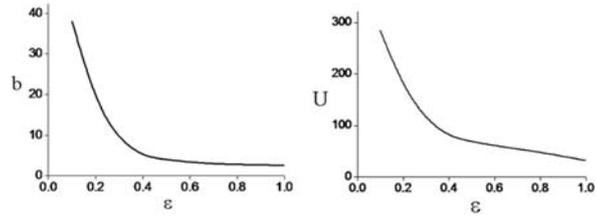

FIG.4

FIG.4 The dipole-mode solitons are stable above the curves in the two parameter planes. All quantities are plotted in arbitrary dimensionless units.

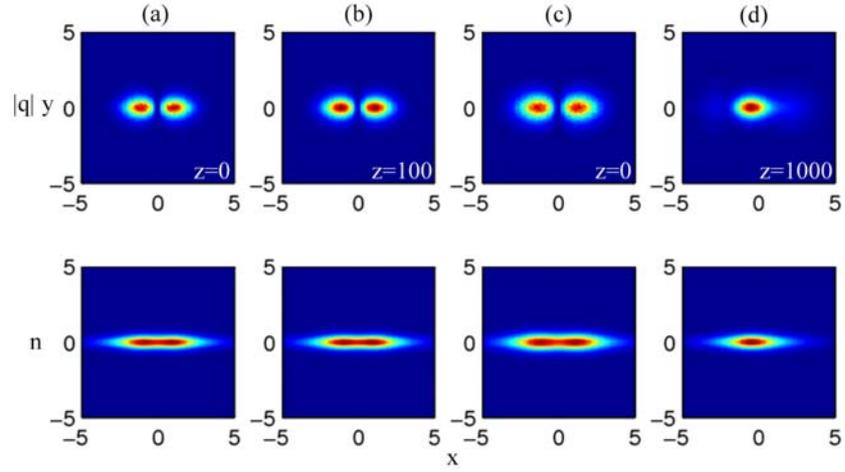

FIG.5

FIG.5 (Color online) Typical examples of the perturbed propagation of stable [columns (a) and (b)] and unstable [columns (c) and (d)] DM solitons, shown in terms of profiles of the absolute value of the field, $|q(x,y)|$ (top row), and the associated nonlinear refractive index, $n(x,y)$ (bottom row). In (a) and (b), $b=3, \varepsilon=1$; in (c) and (d), $b=2, \varepsilon=1$. All quantities are plotted in arbitrary dimensionless units.